# Suspended AlGaAs-Integrated Nonlinear Photonics

*Yuqian Zhang, Changzheng Sun\*, Bing Xiong, Jian Wang, Zhibiao Hao, Lai Wang, Yanjun Han, Hongtao Li, Yi Luo*

Beijing National Research Centre for Information Science and Technology (BNRist), Department of Electronic Engineering, Tsinghua University, Beijing 100084, China
E-mail: czsun@tsinghua.edu.cn

**Abstract**

AlGaAs is a promising integrated nonlinear photonics material with enormous optical nonlinearity and high refractive index. Nevertheless, presently AlGaAs microring resonators exhibiting high quality factors and tight optical confinement rely predominantly on wafer bonding techniques, which entail an intricate fabrication process. Here, we present suspended AlGaAs waveguides and resonators as a viable platform for high efficiency integrated nonlinear photonics. The suspended microring resonator formed by combined plasma dry etching and chemical wet etching exhibits an intrinsic quality factor $Q$ of $2.1\times10^6$, and Kerr comb generation with milliwatt level threshold is recorded. Our demonstration opens up a new prospect for employing AlGaAs devices in integrated nonlinear photonics.

## Introduction

Integrated photonics has the advantages of compact size and low power consumption by integrating optoelectronic devices on the same chip. Over the past few years, significant progress has been made in the field of integrated photonics. One of the key research directions has been the incorporation of nonlinear optical processes into integrated photonic devices, enabling functionalities such as photon generation and manipulation.[1-3] An integrated nonlinear platform with high optical nonlinearity, tight optical confinement, and low propagation loss is essential for the realization of efficient nonlinear photonic devices. So far, nonlinear photonic devices have been implemented on various integrated platforms, such as silicon nitride ($Si_3N_4$),[4-6] silica ($SiO_2$),[7,8] aluminum nitride (AlN),[9-12] gallium nitride (GaN),[13] and lithium niobate ($LiNbO_3$).[14] These integrated platforms have enabled the demonstration of various nonlinear optical effects, including second harmonic generation (SHG),[9] optical parametric oscillation (OPO),[15] Kerr comb generation,[16] and stimulated Raman scattering (SRS).[17]

AlGaAs is considered a promising nonlinear optical material. Firstly, $Al_xGa_{1-x}As$ maintains lattice matched to GaAs substrate throughout the entire Al composition range of $0 \leq x \leq 1$, which allows the epitaxial growth of high crystalline quality AlGaAs/GaAs heterostructures with varying Al concentrations.[18] In particular, two photon absorption (TPA) at telecommunication wavelengths can be avoided by tuning the Al concentration. Secondly, AlGaAs exhibits both a wide transparency window from near- to mid-infrared[19,20] and an enormous intrinsic third order Kerr nonlinearity $n_2 \sim 10^{-17}$ m$^2$W$^{-1}$[21]. Thirdly, AlGaAs has a high refractive index (approximately 3.29 for Al$_{0.2}$Ga$_{0.8}$As at 1550 nm), which helps secure tight optical confinement due to high refractive index contrast against the SiO$_2$ cladding. In addition, AlGaAs also exhibits a high photoelastic coefficient,[22] making it an ideal material for implementing on-chip stimulated Brillouin scattering (SBS).[23-25]

In recent years, extensive endeavors have been undertaken to improve the fabrication of AlGaAs waveguides, leading to remarkable advancements. Minhao Pu *et al.* demonstrated high-quality-factor ($Q > 10^5$) AlGaAs-on-insulator (AlGaAsOI) microrings by employing benzocyclobutene (BCB) as the bonding layer.[26,27] L. Chang *et al.* reported AlGaAsOI microrings with an intrinsic quality factor exceeding $10^6$ formed by direct wafer bonding,[28-30] and sub-milliwatt level Kerr frequency comb threshold was recorded. J. Chiles *et al.* successfully attained a mid-infrared nonlinear photonics platform by means of suspended AlGaAs on silicon.[31] Suspended AlGaAs or GaAs devices with air cladding have also been demonstrated on GaAs substrates.[32,33] However, to date, suspended AlGaAs waveguides or microresonators with low propagation losses and high $Q$ factors remain a challenge.

In this work, we present an air-clad AlGaAs nonlinear platform formed on GaAs substrate, which allows tight optical confinement and low propagation loss. The fabricated AlGaAs microring resonators exhibit an intrinsic $Q$ value of up to $2.1 \times 10^6$, corresponding to a transmission loss as low as 0.28 dB/cm, and milliwatt-level Kerr comb threshold is demonstrated. The proposed scheme can be implemented with a relatively simple fabrication process, as it eliminates the need for wafer bonding. Unlike the AlGaAsOI platform,[26-30,34] the suspended AlGaAs waveguide does not involve SiO$_2$ or Al$_2$O$_3$ claddings, making it suitable for spectroscopy applications in the mid-infrared.[35] Moreover, the suspended waveguide structure allows simultaneous confinement of both optical and sound waves, which is advantageous for the realization of Brillouin integrated photonics.[2]

## Device Design and Fabrication

**Figure 1**a-c illustrate the proposed device structure. The microring is formed on an Al$_{0.2}$Ga$_{0.8}$As waveguide layer to avoid TPA while maintaining high second- and third-order nonlinear coefficients. Suspended waveguide is formed by removing the underlying Al$_{0.8}$Ga$_{0.2}$As sacrificial layer. The large refractive index contrast between AlGaAs and air allows tight optical confinement, thus reducing the mode area and effectively enhancing nonlinearity processes in the suspended waveguide. Furthermore, air claddings help achieve anomalous dispersion around 1550 nm for the suspended AlGaAs microring, which is crucial for Kerr frequency comb generation.[16] In this work, the

Al$_{0.2}$Ga$_{0.8}$As waveguide layer is set to be 600 nm, and the group velocity dispersion (GVD) curves for different ridge waveguide widths are shown in Figure 1d, assuming an etching depth of 500 nm. Anomalous dispersion is attainable for both TE$_{00}$ and TM$_{00}$ modes with a narrow ridge width of 600 nm. Figure 1e shows the simulated optical field distribution of TM$_{00}$ mode in a 600-nm-wide AlGaAs waveguide. Thanks to its high refractive index and strong optical confinement, the mode area of the suspended AlGaAs waveguide is about 0.30 μm$^2$ for TM$_{00}$ mode or 0.31 μm$^2$ for TE$_{00}$ mode, which is much smaller than that of a Si$_3$N$_4$ waveguide commonly adopted for comb generation. Consequently, enhanced nonlinear processes are expected in suspended AlGaAs waveguides. The width of the bus waveguide is taken to be 520 nm to ensure single-mode operation, and pulley coupling scheme is adopted to implement efficient coupling to the microring.

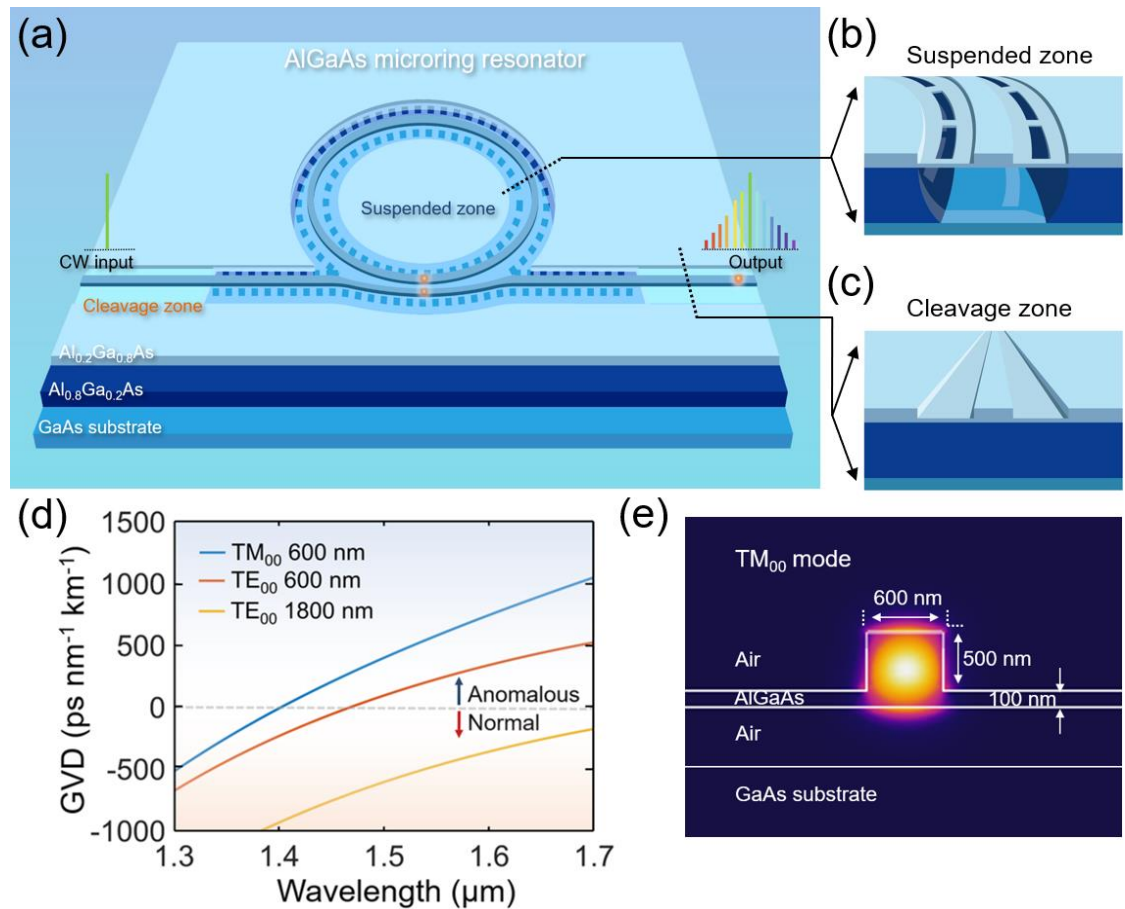

**Figure 1**. Suspended AlGaAs nonlinear platform. a) Schematic of the suspended AlGaAs microring resonator. Localized enlargement of the waveguide cross-section in b) the suspended zone and c) the cleavage zone. d) Group velocity dispersion (GVD) of the fundamental TM/TE modes for partially etched AlGaAs waveguides with different widths. The etching depth is 500 nm. e) Light field distribution of the TM$_{00}$ mode at 1550 nm, indicating effective light confinement within the suspended AlGaAs microring.

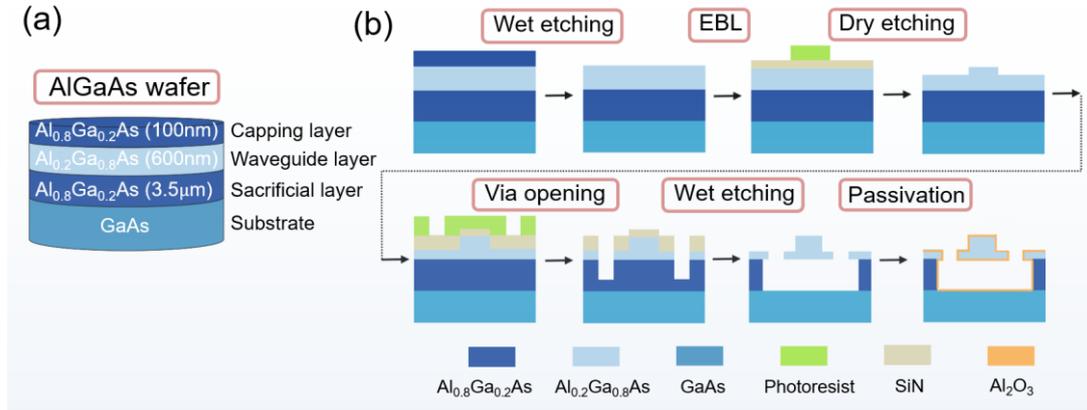

**Figure 2.** a) Epitaxial structure of the device. b) Fabrication processes of the suspended microcavity.

As shown in **Figure 2**a, the wafer used for device fabrication consists of three AlGaAs layers with varying Al components grown on [001] GaAs substrate by metal-organic chemical vapor deposition (MOCVD). A 600-nm-thick $Al_{0.2}Ga_{0.8}As$ layer serves as the waveguide layer. An underlying 3.5-μm-thick $Al_{0.8}Ga_{0.2}As$ is employed as the sacrificial layer, whose thickness is so chosen as to ensure negligible light leakage to the GaAs substrate, while keeping the etching time for the sacrificial layer removal at an acceptable level. The wafer is covered with a 100-nm-thick $Al_{0.8}Ga_{0.2}As$ capping layer, so as to prevent oxidation of the waveguide layer. The high aluminum content capping and sacrificial layers can be selectively removed by chemical etching.

The fabrication processes of the suspended AlGaAs microrings are shown in Figure 2b. Firstly, the waveguide layer is exposed by removing the capping layer in 2% diluted HF for 60 seconds, and then successively cleaned for 30 seconds in HCl (0.3 mol/L) and $NH_4OH$ (1.3 mol/L). Diluted HF exhibits high etch rate for AlGaAs with high Al content, making it an ideal selective etchant.[36] However, after the HF treatment, insoluble byproducts such as $AlF_3$ tend to increase the surface roughness.[37] In Ref. [37], InAlP sacrificial layer was adopted instead of AlAs to avoid the formation of insoluble residuals during the HF etching. In this work, $Al_{0.8}Ga_{0.2}As$ capping and sacrificial layers are employed, so as to simplify the epitaxial growth.[30] By using diluted HCl, the residuals can be dissolved and form highly soluble halogens, which can then be effectively removed by $NH_4OH$. To ensure the complete removal of the residual fluorides, the wafer is further cleaned in *N*-methyl-2-pyrrolidone at 120°C for 10 minutes.[38] **Figure 3**a plots the surface morphology of the $Al_{0.8}Ga_{0.2}As$ capping layer characterized by atomic force microscopy (AFM), indicating a root-mean-square (rms) roughness of 0.36 nm over a 5×5 μm$^2$ region for the as-grown surface. After capping layer removal in diluted HF, an rms roughness of 1.78 nm is recorded for the waveguide layer surface, due to the presence of residual etching byproducts, as shown in Figure 3b. As revealed by Figure 3c, further treatment in hydrochloric acid, ammonia water, and *N*-methyl-2-pyrrolidone results in a surface roughness of 0.31 nm, which is on par with the as-grown surface roughness, indicating the effectiveness of our cleaning procedure.

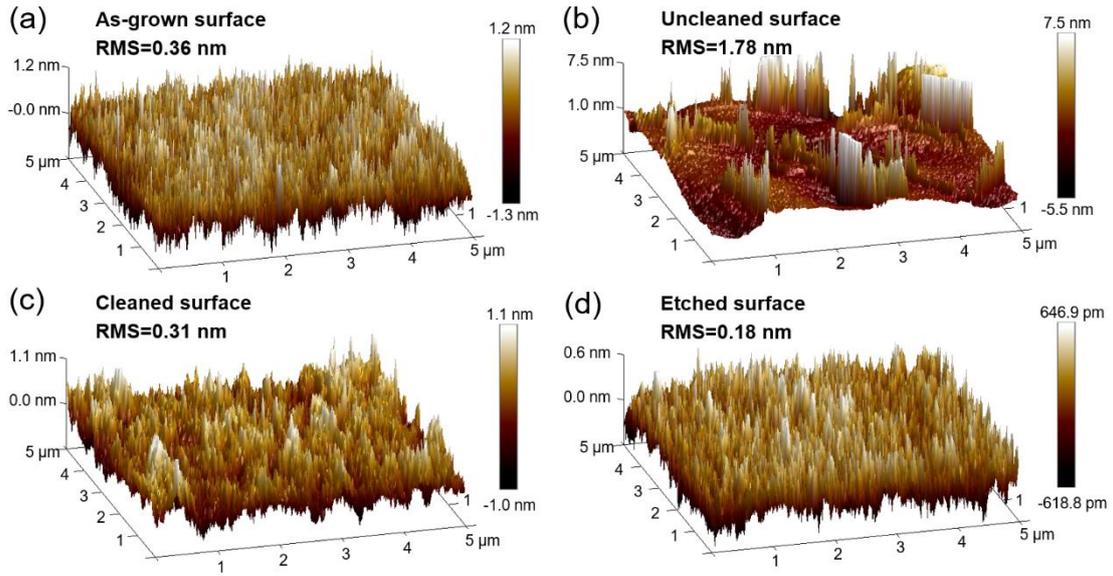

**Figure 3**. AFM images of the AlGaAs wafer surface after different process steps. a) As-grown $Al_{0.8}Ga_{0.2}As$ capping layer; b) $Al_{0.2}Ga_{0.8}As$ waveguide layer after capping layer removal in diluted HF; c) $Al_{0.2}Ga_{0.8}As$ waveguide layer after treatment in diluted hydrochloric acid, ammonia water, and *N*-methyl-2-pyrrolidone; d) dry-etched surface of $Al_{0.2}Ga_{0.8}As$ waveguide layer. The rms roughness in a-d) is 0.36, 1.78, 0.31 and 0.18 nm over a 5×5 μm² region, respectively.

A 150-nm-thick $SiN_x$ layer is then deposited on top of the wafer by plasma enhanced chemical vapor deposition (PECVD) as the hard mask for subsequent dry etching. The microrings and associate bus waveguides are patterned by electron beam lithography (EBL) with ZEP520A photoresist. The pattern is transferred to the $SiN_x$ hard mask by inductively coupled plasma (ICP) dry etching with $SF_6$ and $O_2$ mixture.

The significant electron back-scattering of the AlGaAs layers and the GaAs substrate often result in prickly edges of the ZEP520A photoresist after EBL, which in turn leads to rough AlGaAs sidewalls after dry etching, thus negatively affecting the *Q* value of the fabricated microrings. Thermal reflow of the photoresist[39] is carried out to address this issue, taking advantage of its strong temperature sensitivity and flowability at specific temperatures. In our experiment, the sample is heated on a hot plate at 150°C for 4 minutes to induce melting and self-healing of the photoresist mask. As revealed by the scanning electron microscope (SEM) images shown in **Figure 4**a,b, the edge smoothness of the AlGaAs waveguide improves remarkably after the thermal reflow treatment.

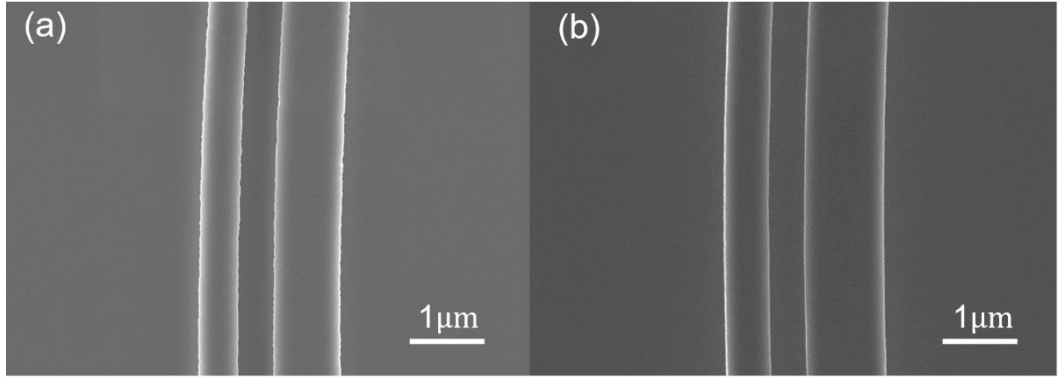

**Figure 4**. SEM images of etched AlGaAs waveguides a) without and b) with thermal reflow treatment. The images are taken after removing the ZEP520A photoresist and the SiN$_x$ hard mask on top of the waveguides.

Dry etching of AlGaAs is typically carried out with Cl-based gases, such as Cl$_2$/N$_2$,[28] BCl$_3$/Ar,[31] and SiCl$_4$/Ar.[32] For suspended AlGaAs waveguides, it is crucial to form partially etched structures with precisely controlled etching depth. BCl$_3$/Ar based ICP dry etching is adopted to ensure a smooth etched surface and an adequate etch rate for precise control of the etching depth. An etch rate of approximately 3.9 nm/s can be secured for a gas flow of BCl$_3$:Ar = 12sccm:18sccm, together with 0.2 Pa chamber pressure, 350 W ICP power, and 90 W RF power. An rms roughness of 0.18 nm over a 5×5 μm$^2$ region for the etched surface is shown in Figure 3d, which is smoother than the as-grown wafer surface shown in Figure 3a. After removing the SiN$_x$ hard mask in 2% diluted HF, the wafer is covered by another 350-nm-thick SiN$_x$ layer for subsequent processing steps.

To form suspended AlGaAs waveguides, via holes are formed on both sides of the microring and the bus waveguides to facilitate the removal of the underlying sacrificial layer. ICP dry etching in Cl$_2$/BCl$_3$/Ar mixture is employed to etch through the waveguide layer and into the sacrificial layer, resulting in a total etching depth of 2.5-3 μm. The addition of Cl$_2$ increases the etch rate to 2.6 μm/min. To minimize the influence of via holes on the *Q* factor of the microring, the edge-to-edge distance between the ring waveguides and the via holes is 2 μm except for the coupling area.

After via hole opening into the sacrificial layer, the device is wet-released in 2% diluted HF for 1 minute, and then placed in deionized water for 5 minutes. This process is repeated for three cycles and the sample is stirred during the entire treatment to ensure timely discharge of the etching byproducts. To eliminate the residual etching byproducts attached to the upper and lower surfaces of the waveguide, which may deteriorate the *Q* factor of the microring, additional cleaning procedures are employed. The cleaning protocol involves the following steps: firstly, immerse the device in HCl (0.3 mol/L) and stir for 90 seconds; then clean the device in NH$_4$OH (1.3 mol/L) for 90 seconds, followed by a treatment in *N*-methyl-2-pyrrolidone at 120°C for 30 minutes. The device is then immersed in isopropanol for 3 minutes. Finally, the device is placed on a 120°C hot plate for 2 minutes to evaporate the isopropanol.

AlGaAs is known to be susceptible to oxidation when exposed to air, resulting in elevated waveguide losses. To mitigate this issue, atomic layer deposition (ALD) is employed to encapsulate the suspended AlGaAs waveguide by a 10-nm-thick aluminum oxide ($Al_2O_3$),[40] thus providing a barrier against oxygen in the ambient environment. Passivation of the AlGaAs waveguide surface also helps reduce the surface absorption due to defect states at material surfaces, thus improving the $Q$ factor of the fabricated device.[37, 38, 40] Meanwhile, according to our simulations, the 10-nm-thick $Al_2O_3$ coating has a negligible influence on the dispersion characteristics of the AlGaAs microring.

For suspended structures, it is of utmost importance to achieve efficient coupling with optical fibers, while simultaneously preserving the structural integrity of the device. Though grating coupling[33,41] is an effective way to avoid potential damage to the suspended waveguide during cleavage, it suffers from relatively high insertion loss. Thanks to the substantial refractive index difference ($\Delta n \approx 0.3$ at 1550 nm) between the $Al_{0.2}Ga_{0.8}As$ waveguide and the $Al_{0.8}Ga_{0.2}As$ sacrificial layer, low light leakage is possible without removing the underneath sacrificial layer. A 300-μm-long cleavage zone is formed on either end of the device for robust facet cleavage, and the sacrificial layer in the cleavage zone remains intact. The width of the bus waveguide in the cleavage zone is tapered to 3 μm for efficient coupling with lensed fibers with a mode field diameter (MFD) of 2.5 μm. The simulated coupling loss is 3.5 dB and 3.8 dB for TE and TM modes, respectively. On the other hand, the insertion loss due to the effective refractive index mismatch between the suspended region and the cleavage zone is calculated to be around 1 dB, while the propagation loss induced by light leakage in the cleavage zone is estimated to be less than 0.1 dB. Consequently, the total coupling loss of the proposed device is estimated to be 4.5-5 dB/facet. The SEM images of the fabricated AlGaAs microring resonator are shown in **Figure 5**a-c. As revealed by Figure 5c, the sidewall angle of the waveguide is close to 90º, which helps ensure precise dimension control.

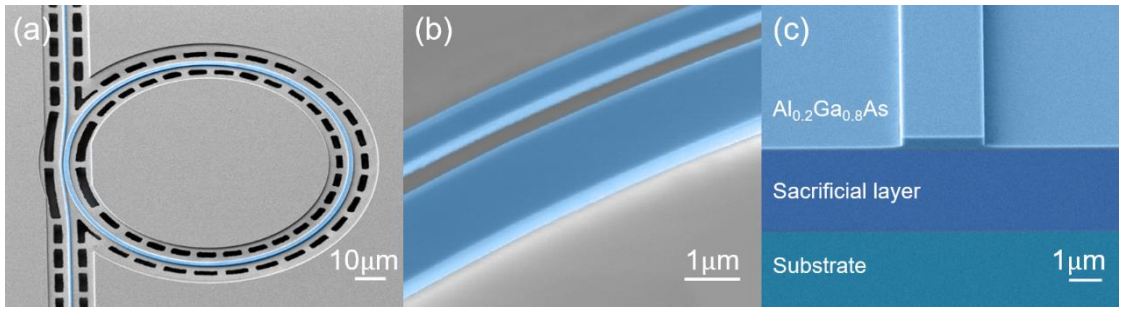

**Figure 5**. SEM images of a) the suspended microring, b) the coupling region, and c) the cleaved bus waveguide facet.

## Optical Measurement

Suspended microrings with the same radius of 50 μm and different waveguide widths are fabricated. Tapered fibers with an MFD of 2.5 μm are employed to measure the transmission spectra shown in **Figure 6**. The insertion loss ranges from 5 to 5.5 dB per facet, in agreement with our theoretical estimation. The intrinsic $Q$ factors of the 600-nm-wide microring are measured to be $6.0\times10^5$ and $5.0\times10^5$ for $TE_{00}$ and $TM_{00}$ modes, corresponding to a waveguide propagation loss of 0.9 dB/cm and 1.1 dB/cm, respectively. On the other hand, the 1800-nm-wide microrings exhibit an intrinsic $Q$ factor of $2.1\times10^6$ and $1.2\times10^6$ for $TE_{00}$ and $TM_{00}$ modes, which are the highest $Q$ values for suspended AlGaAs microrings so far reported. Increasing the microring width reduces the overlap of the optical mode with the waveguide interfaces, thus the 1800-nm-wide multimode waveguide suffers less from sidewall roughness induced scattering loss and exhibits enhanced $Q$ factor.[5]

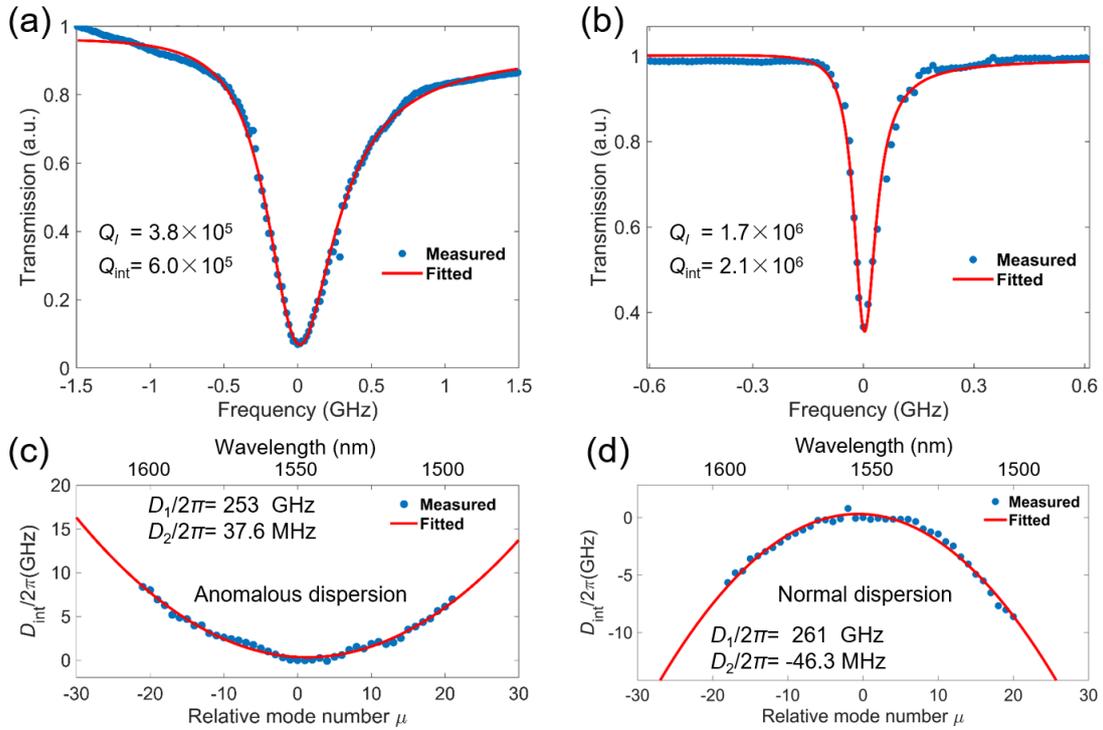

**Figure 6**. Measured resonances of $TE_{00}$ mode for a) 600-nm-wide microring at 1559.05 nm and b) 1800-nm-wide microring at 1550.06 nm. The radius of both microrings is 50 μm. Measured integrated dispersion $D_{int}$ for c) $TM_{00}$ mode of a 600-nm-wide microring and d) $TE_{00}$ mode of a 1800-nm-wide microring.

The dispersion characteristics of the suspended AlGaAs microring is measured with a fiber ring etalon.[42] The resonance $\omega_\mu$ can be expanded as $\omega_\mu = \omega_0 + \mu D_1 + \frac{1}{2}\mu^2 D_2 + \ldots$, where $\mu$ is the relative mode number with respect to the reference resonance $\omega_0$. $D_1/2\pi$ is the free spectral range (FSR) around $\omega_0$, while $D_2/2\pi$ is related to the second-order dispersion. The integrated dispersion $D_{int}(\mu) = \omega_\mu - (\omega_0 + \mu D_1)$ of $TM_{00}$ mode in a 600-nm-wide AlGaAs microring is plotted in Figure 6c. The positive value of the extracted $D_2/2\pi = 37.6$ MHz confirms that the resonator falls in the anomalous dispersion regime. On the other hand, the measured $D_{int}$ of $TE_{00}$ mode in an 1800-nm-wide microring is plotted in Figure 6d. The negative $D_2/2\pi = -46.3$ MHz indicates normal dispersion, in agreement with the simulations shown in Figure 1b.

As a demonstration of the suspended AlGaAs platform for efficient nonlinear photonics applications, Kerr frequency comb generation is carried out. We pump the $TM_{00}$ mode resonance at ~1552 nm ($Q_{int}$ ~ $5.0\times10^5$) in a 600-nm-wide microring, and the output spectra at different on-chip pump powers are recorded, as shown in **Figure 7**a-c. At a pump power of ~ 3.1 mW, parametric oscillation sidebands are observed at 9×FSR away from the pump (Figure 7a). Increasing the pump power to 5.0 mW leads to a significant increase in the number of primary comb lines, and the formation of single-FSR-spaced secondary comb lines can also be observed (Figure 7b). At a pump power of 7.9 mW, a complete comb with single FSR spacing is formed, covering a spectrum range of 1460-1630 nm (Figure 7c).

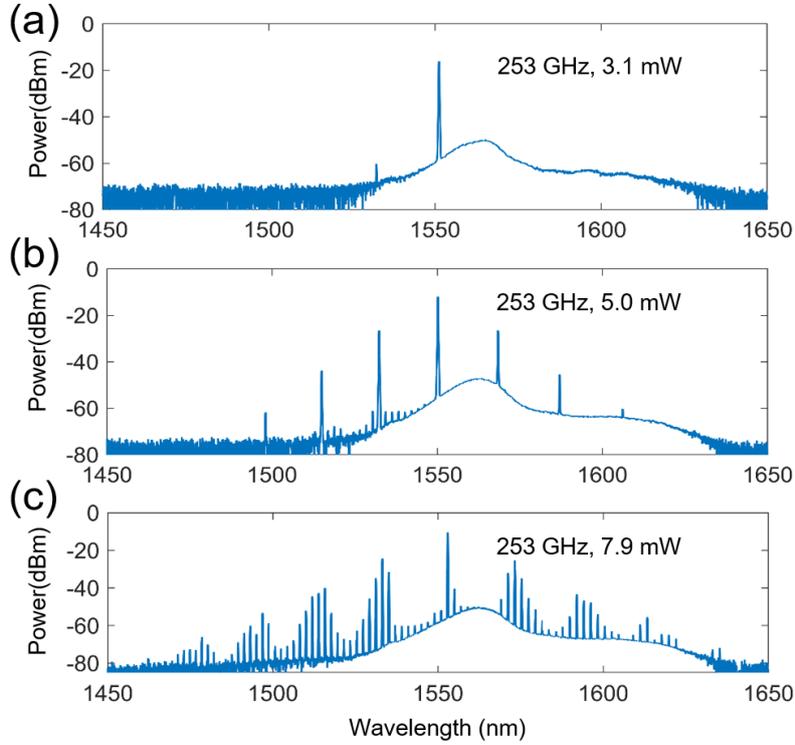

**Figure 7**. Frequency comb spectra from a 600-nm-wide AlGaAs microring with anomalous dispersion, when pumped by TM polarized light with a power of a) 3.1 mW, b) 5.0 mW, and c) 7.9 mW.

Frequency comb generation in suspended AlGaAs microrings with normal dispersion[43] is also recorded. We pump the $TE_{00}$ mode resonance at ~1554 nm ($Q_{int}$~$1.3\times10^6$) in an 1800-nm-wide AlGaAs microring with anomalous dispersion. The variation of the primary OPO sideband power with the on-chip pump power is plotted in **Figure 8**a, indicating a threshold of 1.1 mW. At an on-chip pump power of 6.3 mW, comb lines with 2×FSR spacing are recorded, covering a spectrum range of approximately 1480-1650 nm, as shown in Figure 8b.

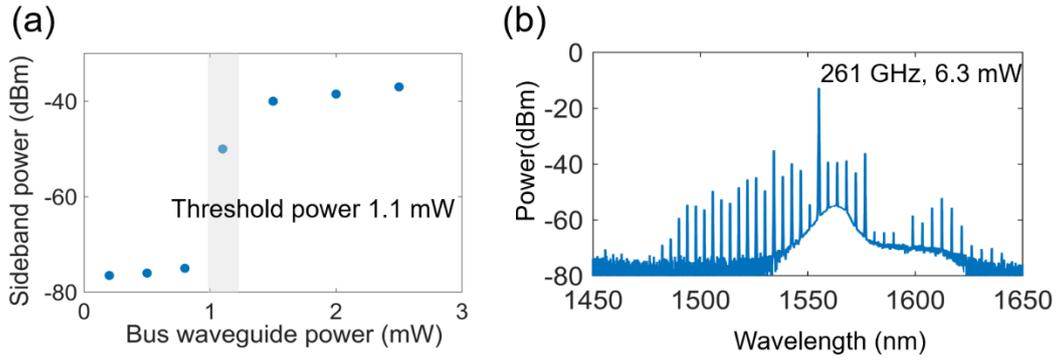

**Figure 8**. a) Measured primary sideband power as a function of pump power in the bus waveguide of a 1800-nm-wide AlGaAs microring. A significant power increase occurs at 1.1 mW. b) Frequency comb spectrum recorded with a TE polarized pump power of 6.3 mW.

## Conclusion

In summary, we have successfully fabricated high-$Q$ suspended AlGaAs microrings. With optimized plasma dry-etching and surface treatment, the intrinsic $Q$ of the fabricated microring is as high as $2.1 \times 10^6$, corresponding to a waveguide loss of 0.28 dB/cm. Thanks to its tight optical confinement, the air clad suspended AlGaAs waveguide exhibits a small mode area of ~ 0.3 µm$^2$, making it an ideal integrated nonlinear photonics platform. Efficient Kerr frequency comb generation with milliwatt level threshold has been demonstrated. The platform can be extended to incorporate second-order nonlinear effects, such as second-harmonic generation. Without the hinderance of $SiO_2$ or $Al_2O_3$ claddings, the air-clad suspended AlGaAs platform lends itself directly to applications in the mid-infrared region. In addition, the suspended structure affords nearly perfect confinement of sound waves and efficient optoacoustic coupling due to the significant acoustic velocity mismatch between the waveguide core and air, thereby making it well-suited for on-chip SBS devices. It is believed that this work will inspire new interest in further development of AlGaAs based integrated nonlinear photonics, e.g. applications in integrated quantum photonics.[44]

## Acknowledgements

This work was supported in part by National Key R&D Program of China (2022YFB2803002); National Natural Science Foundation of China (61975093, 62127814, 62225405, 62235005, 61927811, 61991443, and 61974080); Collaborative Innovation Centre of Solid-State Lighting and Energy-Saving Electronics.